%% file: main.tex
\documentclass{article} %
\usepackage[dvipsnames]{xcolor}
\usepackage[hyphens]{url}

\usepackage[preprint]{neurips_2024}

\input{math_commands.tex}

\usepackage{hyperref}
\usepackage{xspace}
\usepackage{graphicx}
\usepackage{booktabs}
\usepackage{listings}
\usepackage{multirow}
\usepackage{makecell}
\usepackage{amssymb}
\usepackage{makecell}
\usepackage[capitalise]{cleveref}
\usepackage{siunitx}
\usepackage{paralist}
\usepackage{wrapfig}
\usepackage{makecell}

\usepackage{graphicx}
\usepackage{subcaption}

\hypersetup{
    colorlinks,linktoc=all,pagebackref=true,
    linkcolor=BrickRed,
    citecolor=PineGreen,
    filecolor=Mulberry,
    urlcolor=NavyBlue,
    menucolor=BrickRed,
    runcolor=Mulberry,
}
\usepackage{tablefootnote} %

\makeatletter
\DeclareRobustCommand\onedot{\futurelet\@let@token\@onedot}
\def\@onedot{\ifx\@let@token.\else.\null\fi\xspace}

\makeatother

\NewCommandCopy{\ORIcitep}{\citep}
\DeclareRobustCommand{\citep}{\leavevmode\unskip~\ORIcitep}

\NewCommandCopy{\ORIcitet}{\citet}
\DeclareRobustCommand{\citet}{\leavevmode\unskip~\ORIcitet}
\AtBeginEnvironment{appendices}{\crefalias{section}{appendix}}
\AtBeginEnvironment{appendices}{\crefalias{subsection}{appendix}}

\newcommand{\bench}{SWE-PolyBench\xspace}
\newcommand{\benchSmall}{SWE-PolyBench500\xspace}
\newcommand{\SWEBench}{\texttt{SWE}}
\newcommand{\SWEBenchVerified}{\texttt{SWEv}}
\newcommand{\PB}{\texttt{PB}}
\newcommand{\PBsmall}{\texttt{PB500}}

\newcommand{\GTNodes}{\ensuremath{\mathcal{N}}}
\newcommand{\predictedNodes}{\ensuremath{\mathcal{\hat{N}}}}

\newcommand{\agent}{\ensuremath{f_\text{LLM}}}
\newcommand{\problemStatement}{\ensuremath{p}}

\newcommand{\PtoPTestVar}{\ensuremath{t_{\texttt{P2P}}}}
\newcommand{\FtoPTestVar}{\ensuremath{t_{\texttt{F2P}}}}

\newcommand{\meanerror}[2]{\ensuremath{#1}\scriptsize \ensuremath{_{\pm #2}}}

\newcommand{\nodes}{\ensuremath{\mathcal{V}}}
\newcommand{\edges}{\ensuremath{\mathcal{E}}}
\newcommand{\nodeLabels}{\ensuremath{\mathcal{L}}}
\newcommand{\rootNode}{\ensuremath{r}}
\newcommand{\nodeLabelingFunc}{\ensuremath{\lambda}}
\newcommand{\lineSpan}{\ensuremath{S}}
\newcommand{\rangeAssignFunc}{\ensuremath{\sigma}}
\newcommand{\change}{\ensuremath{c}}

\newcommand{\HFLink}{\mbox{https://huggingface.co/datasets/AmazonScience/SWE-PolyBench}}
\newcommand{\RepoLink}{\mbox{https://github.com/amazon-science/SWE-PolyBench}}

\newcommand{\aiderPB}{\texttt{Aider-PB}}
\newcommand{\sweagentPB}{\texttt{SWE-Agent-PB}}
\newcommand{\agentlessPB}{\texttt{Agentless-PB}}

\title{\bench: A multi-language benchmark for repository level evaluation of coding agents}

\author{%
Muhammad Shihab Rashid\thanks{Equal contribution} \quad Christian Bock\footnotemark[1] \quad Yuan Zhuang \quad Alexander Buchholz\thanks{Work done while at AWS}\\
\textbf{Tim Esler}\footnotemark[2] \quad \textbf{Simon Valentin} \quad \textbf{Luca Franceschi} \quad \textbf{Martin Wistuba} \quad \textbf{Prabhu Teja Sivaprasad}\\
\textbf{Woo Jung Kim} \quad \textbf{Anoop Deoras} \quad \textbf{Giovanni Zappella} \quad \textbf{Laurent Callot}\\
AWS AI Labs\\
\texttt{lcallot@amazon.com}\\
}

\begin{document}

\maketitle

\begin{abstract}
  \input{content/abs}

\end{abstract}

\begin{figure}[h!]
    \centering
    \includegraphics[width=0.9\linewidth]{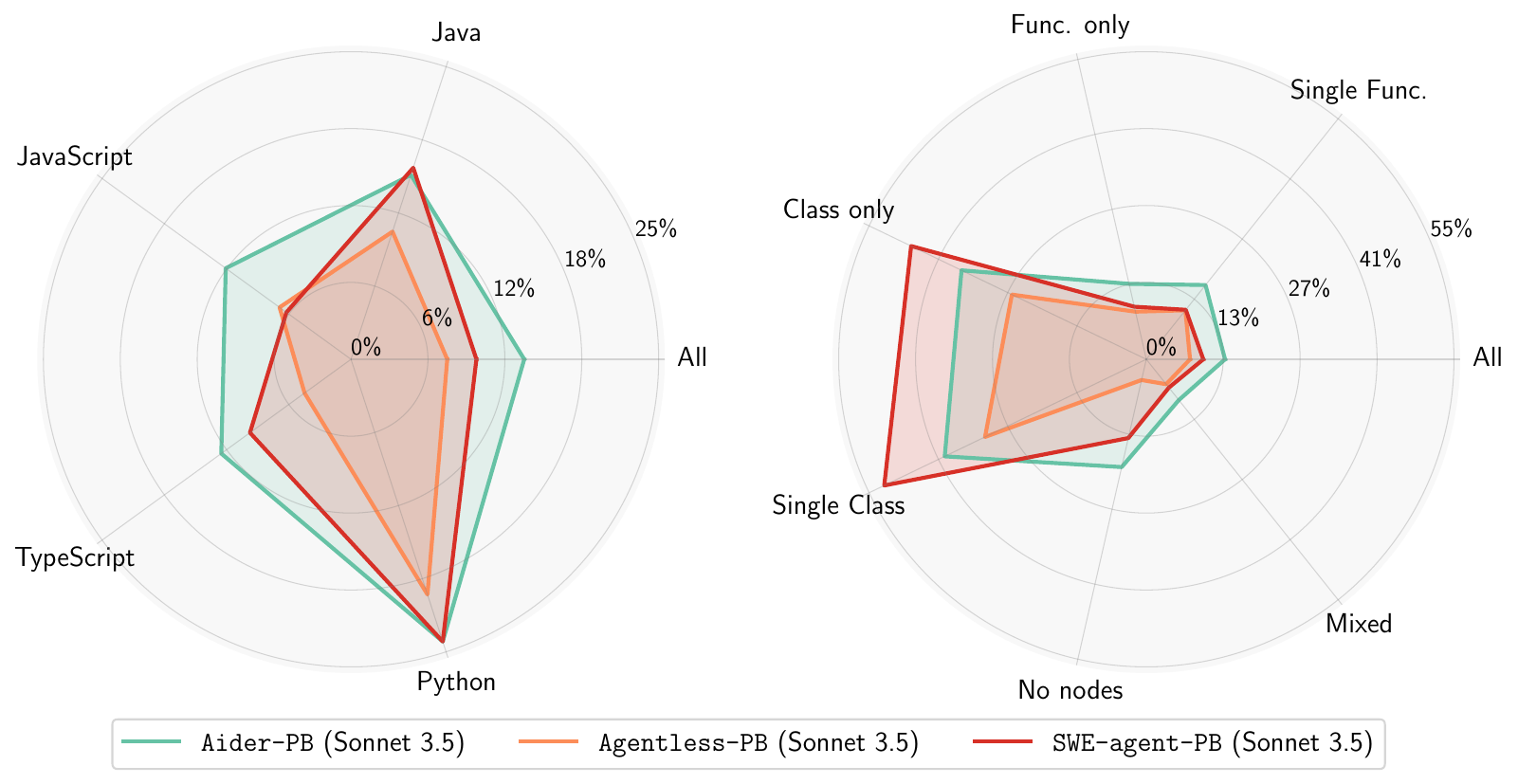}
    \caption{Pass rates of coding agents across programming languages~(left) and across subsets of different complexities based on syntax tree nodes. The right plot categorizes changes by type (class or function) and scope (single or multiple), with "No nodes" indicating no class or function changes and "Mixed" requiring both.}
    \label{fig:radar plots full}
\end{figure}

\section{Introduction}
\input{content/intro}

\section{Related work}
\input{content/related}

\section{Building \bench}
\input{content/data-overview}

\section{\bench Characteristics}\label{sec:characteristics}
\input{content/characteristics}

\section{Evaluating Open-Source Coding Agents}
\input{content/experiments}

\section{Limitations}
\input{content/limitations}

\section{Conclusions}
\input{content/conclusions}

\begin{ack}
We thank Hamed Soleimani, Meysam Feghi, Justice Tomlinson, Adshayan Balendra, Shubham Srivastav, and Andrey Strizhevsky for their valuable contribution in building the dataset and feedback on the evaluation framework.
\end{ack}

\bibliography{main}
\bibliographystyle{plainnat}

\input{content/appendix}

\end{document}

%% file: math_commands.tex
\usepackage{amsmath,amsfonts,bm}

\def\eqref#1{equation~\ref{#1}}

\def\1{\bm{1}}

\DeclareMathAlphabet{\mathsfit}{\encodingdefault}{\sfdefault}{m}{sl}
\SetMathAlphabet{\mathsfit}{bold}{\encodingdefault}{\sfdefault}{bx}{n}

%% file: content/abs.tex
Coding agents powered by large language models have shown impressive capabilities in software engineering tasks, but evaluating their performance across diverse programming languages and real-world scenarios remains challenging.
We introduce \bench, a new multi-language benchmark for repository-level, execution-based evaluation of coding agents.
\bench contains \num{2110} instances from 21 repositories and includes tasks in Java~(165), JavaScript~(1017), TypeScript~(729) and Python~(199), covering bug fixes, feature additions, and code refactoring.
We provide a task and repository-stratified subsample~(\benchSmall) and release an evaluation harness allowing for fully automated evaluation.
To enable a more comprehensive comparison of coding agents, this work also presents a novel set of metrics rooted in syntax tree analysis.
We evaluate leading open-source coding agents on \bench, revealing their strengths and limitations across languages, task types, and complexity classes.
Our experiments show that current agents exhibit uneven performances across languages and struggle with complex problems while showing higher performance on simpler tasks.
\bench aims to drive progress in developing more versatile and robust AI coding assistants for real-world software engineering.
Our datasets and code are available at: \RepoLink

%% file: content/intro.tex
Coding agents are language model-based, autonomous systems able to create or modify software with limited human inputs.
Over the last year, coding agents have garnered substantial attention due to their potential to dramatically enhance human productivity.
The current generation of coding agents exhibit impressive performance on a wide-range of
text-based tasks like code
completion\citep{guo2023longcoder,ding-etal-2024-cocomic}, code
translation\citep{szafraniec2023code}, documentation generation\citep{luo2024repoagent}, unit test
generation\citep{alshahwan2024automated}, debugging\citep{tian-etal-2024-debugbench}, and
conversational code generation\citep{nijkamp2023codegen}.
At the same time, the creation of efficient and effective agents is still an
open research problem and their effectiveness in different scenarios is far
from being understood.
This led to a recent explosion in benchmark creation to assess the coding effectiveness of said systems in controlled environments.
In particular, SWE-Bench~\citep{jimenez2024swebench} which measures the performance of systems in the context of GitHub issues has spurred the development of capable coding agents resulting in over 50 leaderboard submissions, thereby becoming the de-facto standard for coding agent benchmarking.

Despite its significant impact as a pioneering benchmark, SWE-Bench, and in particular its ``verified'' subset~\citep{sweverified}, also shows some limitations.
It contains only Python repositories, the majority of tasks are bug fixes, and with over \SI{45}{\percent} of all tasks, the \verb|Django| repository is significantly over-represented.
Lastly, optimizing for a single dataset results in ``overfitted'' agents whose capabilities no longer generalize, a phenomenon reminiscent of Goodhart's Law~\citep{goodhart1976monetary}.
If the community continues to focus primarily on benchmarks with narrow scope, we risk stunting the development of truly versatile and robust AI coding assistants.

To address these gaps, we are releasing a new family of datasets that aims
to provide a diverse benchmarking
environment for AI coding agents.
Our datasets comprise pull requests~(PRs) and issues from 21 repositories from four of the ten most popular languages according to~\citet{stackoverflow2023}, and three different task categories.
In designing this family of datasets, we strived to strike a good balance
between comprehensiveness and practicality.
We classified tasks into categories~(feature request, bug fix, and refactoring task) to help
researchers quantifying the strengths and weaknesses of their agents.
Finally, we also derive \benchSmall, a representative subset to allow developers and scientists with limited resources to iterate on experiments in an efficient manner.

To enhance the evaluation framework, we introduce novel metrics based on syntax tree analysis: specifically, $\texttt{file}$ and $\texttt{node}$ retrieval metrics. While the existing $\texttt{resolved\_rate}$ serves as the primary performance indicator for coding agents, it provides limited insight into an agent's ability to comprehend and navigate complex codebases. SWE-PolyBench addresses this limitation by incorporating new metrics based on Concrete Syntax Tree (CST) node analysis, complementing the traditional file-level localization measurements. As shown in Figure~\ref{fig:radar plots full}, leading open-source agents demonstrate varying performance across different programming languages, with notable difficulties in tasks requiring complex, multi-file modifications or extensive code changes.

%% file: content/related.tex
\label{sec:related}

Code generation has a long history extending back to program
synthesis\citep{10.1145/362566.362568,gulwani2017program}. Here we focus on the recent work using
large language models~(LLM) for code generation\citep{jiang2024survey} and their corresponding
methods for evaluation.

Prior benchmarks for LLM based code generation can be broadly categorised into
retrieval-free and retrieval-augmented, based on whether retrieval of salient snippets for editing is
required prior to code generation.
Several retrieval-free benchmarks have been proposed like
APPS\citep{hendrycks2021measuring}, HumanEval\citep{chen2021evaluating},
MBPP\citep{austin2021program,athiwaratkun2023multilingual}, and MCoNaLa\citep{wang2022mconala}.
Library specific datasets like NumpyEval~\citep{zan2022cert} have been proposed to study
LLMs when fine-tuned on those libraries' APIs.
These benchmarks, providing all necessary information in the prompt, fail to simulate real-world software engineering tasks where identifying the location for edits is as crucial as determining the edits themselves.

Benchmarks can further be classified as execution-free or execution-based, depending on how code correctness is assessed.
Execution-based benchmarks, more common among the aforementioned ones, use predefined tests to verify generated code.

SWE-bench\citep{jimenez2024swebench} is the first dataset that mirrors real-world software
engineering, in which a code generation system is provided with a codebase and a problem statement,
and is tasked to edit that codebase to solve the problem.
Similar to \bench, SWE-bench was collected
from GitHub issues of popular Python repositories.
This dataset is execution-based, and the tests
for correctness are the unit tests automatically extracted from the repositories themselves.
SWE-bench has spurred a wave of code generation models that have gone from $0.17\%$ to $33.83\%$
pass rate within one and a half years.
SWE-bench Multimodal includes visual elements from front-end
tasks\citep{yang2024swebenchmultimodalaisystems}. SWE-Lancer\citep{miserendino2025swe} extends the
kinds of problems to ones where the task is to choose between multiple code implementations. These
datasets are both execution-based and retrieval-augmented, and thus have been of particular
interest. Extending the topic of the problems being considered,
Baxbench\citep{vero2025baxbenchllmsgeneratecorrect} is security focussed and shows that most LLMs
generate insecure code.

Execution-free benchmarks are in their nascency. Agent-as-a-judge\citep{zhuge2024agent} uses LLM
agents to measure the performance of other agents based on the requirements specified. While this
reduces the costs associated with collecting verification tests, an LLM judgements may not be as
reliable as unit testing.

Most of the discussed benchmarks present several limitations.
A large majority of the benchmarks
focus on Python for bug-fixing problems, with the exception of
SWE-bench-java~\citep{zan2024swebenchjava} and SWE-bench Multimodal~\citep{yang2024swebenchmultimodalaisystems}.
However, the first contains a relative small number~(91) of instances dominated by \SI{85}{\percent} of \texttt{Jackson}-related issues, with the latter focusing on multi-modality.

%% file: content/data-overview.tex
In this section, we summarize the construction process of \bench and \benchSmall and give an overview of their characteristics.
\bench contains \num{2110} samples covering \num{1572} bug reports, \num{463} feature requests, and \num{62} refactoring request.
The dataset spans four languages, and covers \num{21} different repositories.
\benchSmall is a stratified-sampled version of \bench preserving all repositories and the same task category distribution.
The datasets are publicly available through Hugging Face at \hyperlink{\HFLink}{\HFLink}.
The evaluation harness can be found at \hyperlink{\RepoLink}{\RepoLink}.

\begin{figure}[t]
    \centering
    \includegraphics[width=\textwidth]{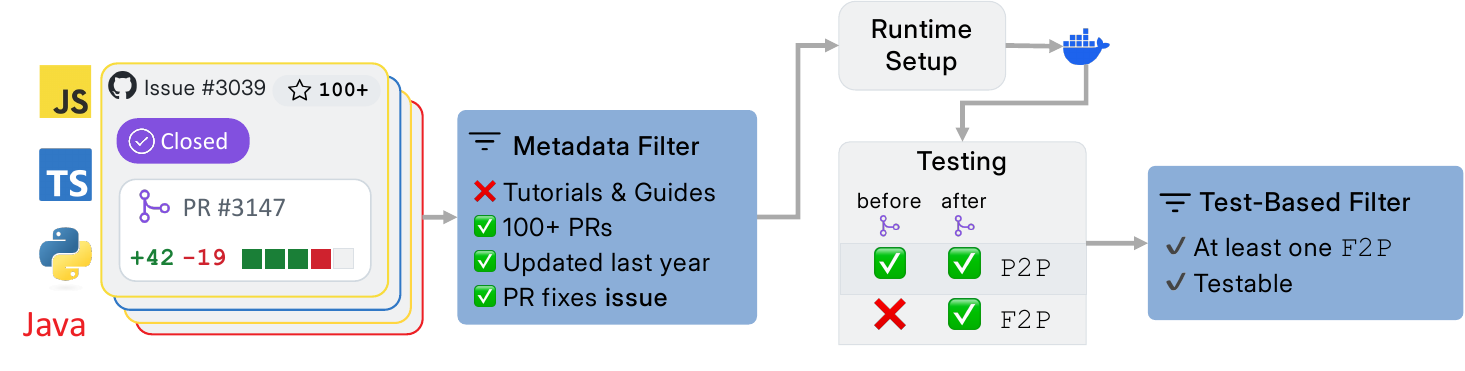}
    \caption{Overview of our dataset generation pipeline.
        We start by collecting pull requests~(PRs) that close an issue from popular repositories across our four target languages.
        After applying a metadata filter, we then create containerized environments for test execution.
        Test outcomes are compared before and after patch application.
        A test is considered a fail-to-pass~(\texttt{F2P}) if it initially fails but passes after applying the patch, a pass-to-pass (\texttt{P2P}) test is defined accordingly.
        The final test-based filter selects PRs that have at least one F2P test and are considered ``testable'' following the definition in \cref{sec:runtime setup}.
        }
        \label{fig:overview}
\end{figure}

\subsection{Data Collection}
\cref{fig:overview} illustrates the full dataset collection pipeline.
We built \bench using real-world GitHub repositories whose primary programming language is either Java, JavaScript~(JS), Python, or TypeScript~(TS) as these represent a significant portion of modern software development projects as per \cite{stackoverflow2023}.
A repository is included if it is implementation-focused~(i.e., excluding guides and tutorials) to focus on real-world code rather than instructional content;
contains at least 100 pull requests~(PRs) to ensure a established history of collaborative development;
was updated in the last 12 months to maintain relevance to current practices;
is permissively licensed to allow unrestricted use for research;
and has English as its primary language to facilitate consistent analysis of code and discussions.
Lastly, a PR is included if it solves an issue and provides respective test code to ensure execution-based verifiability.
This process resulted in 17 repositories for Java, 12 for JS and ten for each of the other languages, yielding a total of
\num{377300}
PRs.
To prevent data contamination and to ensure independent evaluation of model performance, we excluded any repositories found in SWE-Bench from \bench.

\subsection{Runtime Setup}\label{sec:runtime setup}
In addition to the above filtering, we assess whether a PR can be included in the dataset through execution.
In the following we refer to ``test patch'' and ``code patch'' which are defined as follows:
The code patch is the \texttt{git diff} containing changes in the PR that do not relate to the tests.
We refer to this patch also as the ``ground truth patch'' as it is the ground truth code implementation which solves the issue.
The test patch is the \texttt{git diff} containing only changes relating to test updates.
Additionally, the ``base commit'' is the commit onto which the respective PR was merged.

For each PR, to which we also refer to as task, a Docker file is defined to serve as the execution context in which the code base is installed prior to applying any patches.
Given that each programming language has its own package manager, installation procedures, and version requirements, we tailored the setup accordingly.
For example, Java projects commonly use \verb|Maven| for project management and build automation, while JavaScript and TypeScript projects typically rely on \verb|npm| for package and dependency management.
For each repository and/or base commit, we manually configured Docker files to set up the execution environment.
Within this environment, we ran the test suite both before and after applying the code patch.
We then logged two sets of tests: those that transitioned from a `failed' state to a `passed' state (which we refer to as \texttt{F2P} or fail-to-pass), and those that passed both before and after the application of the code patch (which we call \texttt{P2P} or pass-to-pass).

For a PR to be included in the dataset it must contain at least one \texttt{F2P} test.
Lastly, we deem PRs untestable and exclude them if the code patch introduces new files whose contents are tested in the code patch.
This is because created tests cannot reliably evaluate LLM-generated code if functionally correct code was created in unexpected file or function names.
To run execution-based evaluation, all Docker files are published alongside our evaluation harness.

\cref{tab:repo_stats,tab:repo_list} in the appendix provide an overview of the dataset statistics throughout the collection and filtering process.
The average repository size varies significantly across languages, with TypeScript repositories being the largest on average~(8946.0 files) and Python repositories the smallest~(1928.1 files).
Note that these counts include binary and documentation files.

%% file: content/characteristics.tex
\subsection{Contrasting \bench with SWE-Bench}

\begin{table}[t!]
    \centering
    \caption{Contrasting \bench~(\PB) and \benchSmall~(\PBsmall) file statistics and task categories with SWE-Bench~(\SWEBench)\citep{yang2024swe} and SWE-Bench verified~(\SWEBenchVerified)\citep{sweverified}.}
    \begin{tabular}{r|ccccc|cccc}
        \toprule
           & \multicolumn{5}{c}{Modified Files (avg.)} & 
           \multicolumn{4}{c}{Task Category~(\%)} \\
           & Python & Java & JS & TS & All & Bug Fix & Feature Req. & Refac. & Misc. \\
            \midrule
            \SWEBench & \num{1.6} & -- & -- & -- & \num{1.6} &
            \num{75.50} & \num{18.40} & \num{5.14} & \num{0.96} \\
            \SWEBenchVerified & \num{1.2} & -- & -- & -- & \num{1.2} & 
            \num{87.20} & \num{8.60} & \num{4.00} & \num{0.20} \\
            \PB & \num{2.0} & \num{3.6} & \num{2.2} & \num{3.1} & \num{2.6} &
            \num{74.50} & \num{21.94} & \num{2.94} & \num{0.61}  \\
            \PBsmall & \num{2.1} & \num{2.7} & \num{3.6} & \num{1.9} & \num{3.2} &
            \num{51.00} & \num{36.80} & \num{12.20} & \num{0.00} \\
        \bottomrule
    \end{tabular}
    \label{tab:stats files and categories}
\end{table}

We compare key characteristics of \bench~(\PB) and \benchSmall~(\PBsmall), a stratified subset~(see \cref{sec:polybench small} for details), with SWE-Bench~(\SWEBench) and SWE-Bench verified~(\SWEBenchVerified), as the latter two are widely recognized as the current standard for evaluating coding agent performance.

As shown in Table~\ref{tab:stats files and categories}, \bench exhibits higher complexity in terms of modified files across all languages.
Overall \SI{63}{\percent} more files need to be edited to solve a task in \PB\ compared to \SWEBench.
In terms of necessary file modifications, \bench's Java dataset exhibits the highest complexity involving \num{3.6} files on average.
The task categories~(see \cref{sec:task classification} for details on their construction) are similarly distributed between \PB and \SWEBench.
However, in the construction of \PBsmall, we increased the proportion of features requests and refactoring task resulting in \SI{36.80}{\percent} of the former and over \SI{12}{\percent} of the latter, to allow for a more balanced evaluation of a broader range of software engineering scenarios.

The number of modified files can serve as a good proxy for modification complexity but it may not provide the desired level of detail.
\cref{tab:stats node changes} considers change locations in terms of nodes in the concrete syntax tree~(CST) of the modified files~(see \cref{sec:metrics} for details on CST node construction).
In all subsequent analyses, we only consider \texttt{class} and \texttt{function} nodes.
Furthermore, we define changes in a function node not to be counted towards a change in the encompassing class.
More concretely, we only consider the \textit{deepest} node in the CST affected by the code change.

The Node Change Category column in \cref{tab:stats node changes} reflects the percentage of tasks where no \texttt{class} or \texttt{function} node was modified~(None); \textit{only} \texttt{function} nodes are modified~(Func. only); \textit{only} \texttt{class} nodes are modified~(Class only); and \textit{both} node types are modified~(Mixed).
For instance, a ``None'' change might involve modifying configuration files~(e.g., \texttt{.yaml} or \texttt{.json}) or other files that are not typically parsed for class or function structures.
The Node Change Count column provides the average number of node types modified in an instance and the average number of modified nodes~(independently of their type).

\bench exhibits substantial complexity, with Python instances showing \SI{37.69}{\percent} mixed node changes and averaging 5.76 node modifications per instance—notably higher than SWE-bench's \SI{27.99}{\percent} mixed changes and 3.54 modifications.
Java demonstrates the highest complexity among all languages, with \SI{66.06}{\percent} of instances requiring mixed changes and an average of 9.81 node modifications.
Conversely, JavaScript and TypeScript present simpler profiles, with JS having the highest function-only modifications (\SI{84.27}{\percent}) and TS showing the most instances (\SI{30.59}{\percent}) requiring changes not related to classes or functions.
In \benchSmall, these complexity patterns are further amplified, particularly for Java where complexity peaks at \SI{69.60}{\percent} mixed changes and 11.11 average node modifications, reinforcing that \bench presents a structurally complex dataset.

\begin{table}[t!]
    \centering
    \caption{Dataset complexity in terms of changes in concrete syntax tree nodes. Highest numbers per column are highlighted in bold.}
    \begin{tabular}{rc|rrrr|ccc}
        \toprule
           & & \multicolumn{4}{c}{Node Change Category~(\%)}
           & \multicolumn{3}{c}{Node Change Count~(avg.)}\\
           Dataset & Language  & None & \makecell{Func. \\only} & \makecell{Class \\only} & Mixed & Func. & Class & \makecell{Num. \\Nodes} \\
           \midrule
           \SWEBench 
           & Py & \num{1.48} & \num{67.26} & \num{3.27} & \num{27.99} & \num{2.81} & \num{0.72} & \num{3.54} \\
           \SWEBenchVerified 
           & Py & \num{1.80} & \num{77.60} & \num{4.20} & \num{16.40} & \num{1.87} & \num{0.30} & \num{2.18}  \\
           \midrule
           \multirow{4}{*}{\PB} 
           & Py & \num{1.01} & \num{55.78} & \bfseries \num{5.53} & \num{37.69} & \num{4.09} & \num{1.67} & \num{5.76}  \\
           & Java & \num{0.00} & \num{32.12} & \num{1.82} & \num{66.06} & \num{7.35} & \num{2.45} & \num{9.81} \\
           & JS & \num{3.74} & \bfseries \num{84.27} & \num{0.29} & \num{11.70} & \num{2.45} & \num{0.14} & \num{2.60} \\
           & TS & \bfseries \num{30.59} & \num{56.24} & \num{1.78} & \num{11.39} & \num{1.86} & \num{0.21} & \num{2.06} \\
           & All & \num{12.46} & \num{67.82} & \num{1.42} & \num{18.29} & \num{2.78} & \num{0.49} & \num{3.28}\\
           \midrule
           \multirow{4}{*}{\PBsmall}
           & Py & \num{0.00} & \num{52.80} & \num{4.80} & \num{42.40} & \num{4.72} & \num{2.19} & \num{6.91} \\
           & Java & \num{0.00} & \num{28.00} & \num{2.40} & \bfseries \num{69.60} & \bfseries \num{8.42} & \bfseries \num{2.69} & \bfseries \num{11.11} \\
           & JS & \num{6.40} & \num{75.20} & \num{2.40} & \num{16.00} & \num{2.87} & \num{0.19} & \num{3.06} \\
           & TS & \num{28.80} & \num{60.80} & \num{0.80} & \num{9.60} & \num{2.11} & \num{0.14} & \num{2.25} \\
           & All & \num{8.80} & \num{54.20} & \num{2.60} & \num{34.40} & \num{4.54} & \num{1.31} & \num{5.85} \\
        \bottomrule
    \end{tabular}
    \label{tab:stats node changes}
\end{table}

\subsection{LLM-based task classification}\label{sec:task classification}
To gain deeper insights into the dataset, we classify issue descriptions, which we also refer to as problem statements, across two main dimensions.
First, we categorize them based on the nature of the development task, i.e., is it a bug fix, feature request, or a refactoring task.
Second, we evaluate the informativeness of the issue description with respect to the task at hand.
In particular, we provide categories according to the location of needed code changes, hints at the solution, and the overall descriptiveness of the GitHub issue.

\paragraph{Development task classification} Real-world software issues encompass a diverse range of tasks beyond bug fixes. For instance, an issue may involve a refactoring request requiring extensive modifications, or the implementation of a new feature.
A comprehensive evaluation of each task category is essential to identify the strengths and weaknesses of a software agent.
Such analysis is crucial for guiding the research community in the development of agents that are either robust across a variety of tasks or specialized for certain tasks.
To facilitate this, we have categorized \bench's tasks into five distinct categories: Bug Fix, Feature Request, Refactoring, Security, and Testing.
For the classification process, we utilize a large language model (LLM) by providing it with the problem statement and the corresponding ground truth patch, instructing it to classify the task into one of the five predefined categories. The prompt includes a description of these categories, and the complete prompt is detailed in \cref{sec:classification_task}.

\paragraph{Informativeness of problem statements}
As the provided issue descriptions vary according to the type of information they contain, we assess them with respect to the classification initially provided in \cite{xia2024agentlessdemystifyingllmbasedsoftware}.
We describe problem statements along three different dimensions:
\begin{inparaenum}[i)]
  \item how descriptive is the problem statement, 
  \item how much information does the problem statement contain with respect to the desired solution, and
  \item what localization information is available to address the described issue.
\end{inparaenum}
The categories of this annotation are roughly ordered according to the level of informativeness:
the description of the problem ranges from containing sufficient information in natural language~(NL) to none;
hints at the actual solution range from from none to providing the exact solution to providing misleading information;
the localization ranges from being exact in NL to none.
We provide an exact description of these categories in the appendix in \cref{tab:informativeness_problem_statements}. 
Annotating all the problem statements manually is not scalable, therefore we use the help of an LLM 
to obtain the desired classification.
The input to the LLM call are the prompts described in \cref{sec:prompt_classification_ps}, together with the given GitHub issue (i.e., problem statement) and the ground truth patch. For this classification the model does not have access to the actual code repository.

We provide descriptive statistics pertaining to the above classifications in \cref{fig:categories_ps} for \bench and in \cref{fig:categories_ps_sampled} of the appendix for \benchSmall.
Overall we see that the different programming languages exhibit similar categories.
We observe that in the large majority, exact or complete solutions are not provided in the problem statement.
The descriptiveness is dominated by sufficient descriptions in natural language and reproducible examples. 
With respect to the solution content, we see that either no solution or only partial solutions are present in \bench.
The localization of the issue is dominated by exact locations in natural language or keywords.

\begin{figure}[h]
    \centering
    \includegraphics[width=\textwidth]{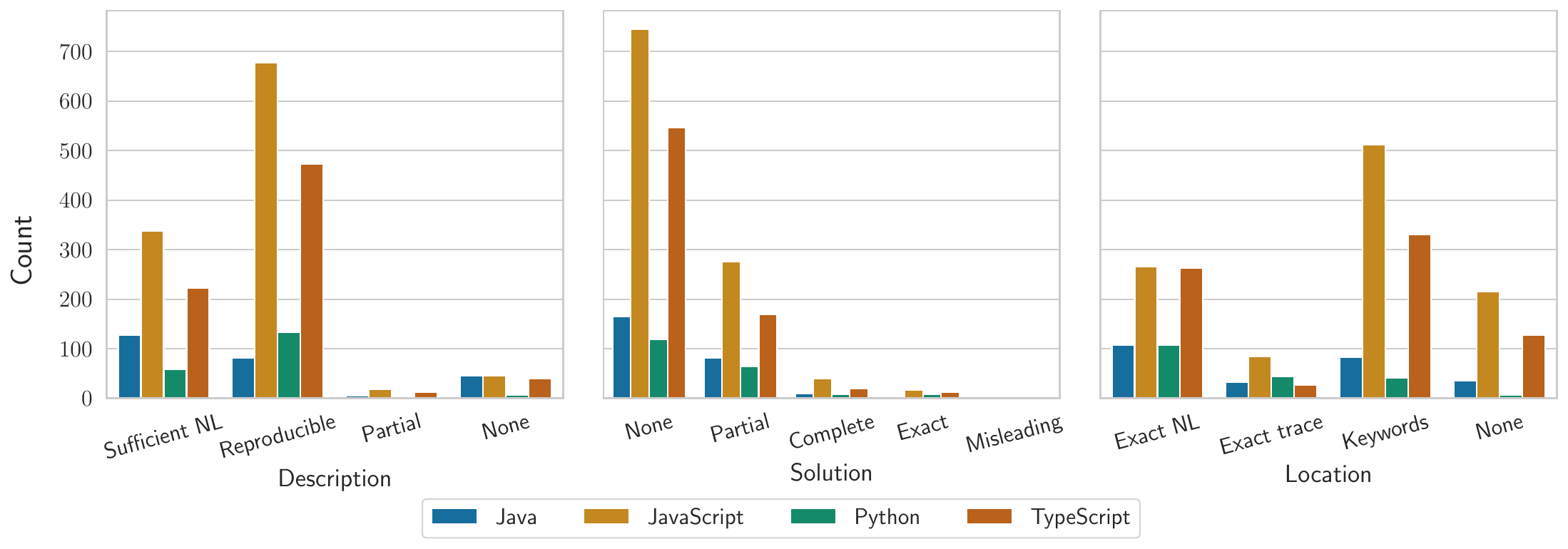}
    \caption{Classification of \bench issue descriptions with respect to their descriptiveness~(left plot), hints at the solution~(middle plot), and information on the localization of the issue~(right plot).
    }\label{fig:categories_ps}
\end{figure}

\subsection{\benchSmall}\label{sec:polybench small}
Following the development task classification, we performed stratified sampling on the full dataset to provide a dataset to the research community of manageable size to experiment with.
Specifically, we selected 500 data points, 125 from each language, with 200 instances each from the bug fix and feature request categories, and 100 instances from the refactoring category.
We also ensured that we have instances from all repositories following same distribution of the full dataset.
We refer to this dataset as \benchSmall~(\PBsmall).

%% file: content/experiments.tex
In this section, we first briefly describe the publicly available coding agents that we evaluated on both \bench and \benchSmall.
We will then detail the technical challenges encountered and the solutions implemented to adapt these methods to our multi-lingual framework.
Lastly, we contrast their performance in terms of pass rate and various retrieval metrics. 

\subsection{Coding Agents}
For our comparison, we selected three open-source agents that are widely recognized and appreciated in both the research community and among practitioners. These include:
\begin{itemize}
    \item \textbf{Aider}~\citep{aider}, an interactive pair programming agent. The agent suggests different changes to the codebase and the user can select or submit their preferences. To run Aider on the benchmarks, we disable the interactive part and always select the agents' suggestions.
    \item \textbf{SWE-agent}~\citep{yang2024swe} which employs an agent-computer interface that can create and edit code files, navigate entire repositories, and execute tests and other programs. 
    \item \textbf{Agentless}~\citep{xia2024agentlessdemystifyingllmbasedsoftware} which uses a three-phase approach to 1) localize, 2) repair, and 3) validate code. It does not rely on autonomous agent-based interactions or complex tools to implement a solution. 
\end{itemize}
We modified and adapted these agents to address the specific challenges of \bench, resulting in their modified versions: \aiderPB, \sweagentPB, and \agentlessPB. All implementations utilize \texttt{Anthropic's Claude 3.5} (claude-3-sonnet-20241022) as the foundation large language model if not stated otherwise. The modifications made to each agent are detailed below.

\subsubsection{Technical Challenges in Making Coding Agents Multi-Lingual}

\paragraph{Aider} During its execution pipeline, Aider~(v0.75.2) includes a validation step that runs preexisting tests against the model-generated patch to ensure it doesn't introduce regressions. This step requires two key components: access to the test execution command and a parser to interpret test results. In the original implementation, these components are specifically tailored for 
Python projects, utilizing \texttt{pytest} as the testing framework. However, adapting this process for \bench presents significant challenges. First, it would require maintaining a comprehensive database of test execution commands for each instance. Second, we would need to develop robust log parsers capable of interpreting test results across diverse testing frameworks. Given these complexities, we opted to exclude this validation step in \aiderPB.

\paragraph{Agentless} 
The original Agentless~(v1.5.0) implementation employs Python-specific tools for its fault localization process, primarily using the \texttt{ast} python module to identify files, functions, and classes requiring modifications, as well as for linting.
This Python-centric approach, however, limits its applicability to other programming languages. Similarly, its bug reproduction mechanism relies on generating and executing Python scripts,
which is not generalizable across different languages. Furthermore, Agentless encounters the same regression testing limitations as Aider.
In our adaptation, \agentlessPB, we address these limitations by incorporating \texttt{tree-sitter} for parsing and extracting code structures across JavaScript, TypeScript, and Java. We also implement language-specific execution commands for bug reproduction scripts. As with \aiderPB, we exclude the regression testing step from the pipeline.

\paragraph{SWE-agent} The original SWE-agent~(v1.0) implementation relies on a containerized environment using \texttt{SWE-ReX} for interacting with repository contents.
While SWE-agent supports custom Docker images, our adaptation process revealed significant compatibility challenges.
It imposes specific requirements on pre-installed packages, including \texttt{Python3.11}, \texttt{SWE-ReX}, and \texttt{pipx}, within the provided Docker images. In our adaptation, \sweagentPB, we initially addressed these issues by directly installing the missing packages in our Docker images, which resolved problems for a subset of instances.
For the remaining cases, we explored an alternative approach: building a new Docker image on top of our provided image with a standalone Python installation.
This method successfully isolates the required packages from the base image for many instances. However, it fails for 129 instances due to version incompatibilities between system libraries.
For example, the new Docker image requires a specific \texttt{glibc} library version while some of our images contained older versions.
For these 129 instances, we eventually used the base image of the Javascript instances as generic image, which has a comprehensive list of pre-installed packages and also meets the package requirement by SWE-agent. We then provided it as a custom docker image for SWE-agent. Among the 129 instances, we were able to obtained predictions for 111 instances. For the rest 18 instances, we treated them as empty predictions in \sweagentPB\ when reporting performance metrics.

\subsection{Metrics}\label{sec:metrics}

We assess coding agents by calculating their pass rates and various retrieval metrics, which we formalize in this section.

Let $\mathcal{X}$ be a space of finite sequences of strings. 
Given a problem statement $\problemStatement\in\mathcal{X}$ 
and the contents of a repository $r\in\mathcal{X}$, an LLM-based agent $\agent$ outputs an updated 
repository $r'=\agent(\problemStatement, r)\in\mathcal{X}$\footnote{For simplicity, we take here the agent to be deterministic. In practice, agents are typically stochastic.}. Furthermore, denote by $r' \smallsetminus r \in \mathcal{X}$ the difference between the input repository and the edited repository.

\paragraph{Pass rate} 
We consider \problemStatement\ to be solved by the agent if the execution of a set of tests on $r'$ is successful. 
In line with previous work \citep{jimenez2024swebench}, the tests comprise a number of pass-to-pass (\texttt{P2P}) and fail-to-pass tests (\texttt{P2P}), as detailed in \cref{sec:runtime setup}.
Formally, 
this means that for each instance of \bench we provide two Boolean functions $\PtoPTestVar:\mathcal{X}\to \{0, 1\}$ and $\FtoPTestVar:\mathcal{X}\to \{0, 1\}$ such that $\PtoPTestVar(r)=1$ and $\FtoPTestVar(r)=0$. 
We let $t= \PtoPTestVar \wedge \FtoPTestVar$ be their conjunction. 
Then, we define the pass rate of an agent \agent\ on a dataset $\mathcal{D}$ as follows: 
\begin{equation}
    \texttt{PassRate}(\agent, \mathcal{D}) = 
    \frac{1}{|\mathcal{D}|} \sum_{(\problemStatement, r, t)\in \mathcal{D}} t(\agent(\problemStatement, r)).
\end{equation}

\paragraph{Retrieval scores.}
While pass rate is a central metric to measure code generation performance, it might fail to capture an agent's ability to successfully navigate a repository and localize relevant code elements.
For this reason, we calculate file-level retrieval metrics~(recall and precision) and introduce a new set of CST node-level retrieval metrics which we define below.

Let $r^*\in\mathcal{X}$ be the repository patched with the ground truth patch,
and let $\texttt{F}:\mathcal{X}\to 2^\mathcal{X}$ be a set-valued function that given a diff string, extracts
the file paths changed in the diff.
Then, to compute file-level retrieval scores~(e.g., recall and precision) we use $\texttt{F}(r^*\smallsetminus r)$ as the ground truth set and $\texttt{F}(r'\smallsetminus r)$ as the predicted set.

For node-level retrieval scores, we construct the CST for each changed file of a given \texttt{diff} string.
A CST is a detailed tree representation of the source code that preserves syntactic details while representing structural elements like functions and classes as nodes in the tree.
We then retrieve the modified nodes~(i.e., \texttt{module}, \texttt{class}, \texttt{function}) accounting for their depth in the tree.
More formally, let $\texttt{CST}:\mathcal{X} \times \mathcal{X} \to \mathcal{T}$ be the parsing function that 
given a \texttt{diff} string and a repository, returns a labelled tree representing the portion of the CST touched by the \texttt{diff}.
The labels of each node uniquely identify a section of the repository~(e.g. \texttt{class}, or \texttt{function}), but do not include their actual content.
Let $\texttt{Paths}:\mathcal{T}\to 2^\mathcal{T}$ be a function that extract all root-to-leaf paths of a tree, yielding a set of paths (i.e., trees with degree at most $1$).
Then, to compute node-level retrieval scores we consider
$\texttt{Paths}(\texttt{CST}(r^* \smallsetminus r, r^*))$ to be the ground truth set and
$\texttt{Paths}(\texttt{CST}(r' \smallsetminus r, r'))$ to be the predicted set.
For details about CST construction, please refer to \cref{sec:cst details}.

\begin{table}[t!]
    \centering
    \caption{Pass rates of open source agents on \bench.}
    \small
    \begin{tabular}{cr|cccc|c}
        \toprule
           & & \multicolumn{4}{c}{Language}
           \\
           Agent & Base LLM & Java & JS & TS & Python & Overall
           \\
        \midrule
        \agentlessPB
        & Sonnet 3.5 & \meanerror{10.9}{2.44} & \meanerror{7.2}{0.81} & \meanerror{4.7}{0.77} & \meanerror{20.1}{2.95} & \meanerror{7.8}{0.59} \\
        \midrule
        \sweagentPB
        & Sonnet 3.5 & $\mathbf{\meanerror{16.4}{2.88}}$ & \meanerror{6.5}{0.77} & \meanerror{10.2}{1.10} & $\mathbf{\meanerror{24.1}{3.05}}$ & \meanerror{10.2}{0.66}
        \\
        \midrule
        \multirow{5}{*}{\aiderPB}
        & Sonnet 3.5 
        & \meanerror{15.8}{2.86} & $\mathbf{\meanerror{12.6}{1.04}}$ & $\mathbf{\meanerror{13.0}{1.24}}$ & $\mathbf{\meanerror{24.1}{3.05}}$ & $\mathbf{\meanerror{14.1}{0.77}}$
        \\
        & Deepseek R1
        & \meanerror{12.1}{2.55} & \meanerror{10.1}{0.95} & \meanerror{11.5}{1.17} & \meanerror{18.1}{2.81} & \meanerror{11.5}{0.71} \\
        & Haiku 
        & \meanerror{11.5}{2.49} & \meanerror{8.1}{0.85} & \meanerror{9.7}{1.09} & \meanerror{18.1}{2.81} & \meanerror{9.9}{0.65} 
        \\
        & Mistral-Large 
        & \meanerror{6.7}{1.96} & \meanerror{4.8}{0.66} & \meanerror{6.9}{0.93} & \meanerror{7.0}{1.83} & \meanerror{5.9}{0.51}
        \\
        & Llama 3.3 70B 
        & \meanerror{9.1}{2.24} & \meanerror{4.2}{0.62} & \meanerror{6.4}{0.91} & \meanerror{11.1}{2.27} & \meanerror{6.0}{0.52} 
        \\
        & \makecell[r]{DeepSeek-R1-\\Distill-Llama-70B} & \meanerror{5.5}{1.78} & \meanerror{3.5}{0.57} & \meanerror{5.8}{0.86} & \meanerror{12.6}{2.40} & \meanerror{5.3}{0.48}
        \\
        \bottomrule
    \end{tabular}
    \label{tab:agent-results-pass-rate}
\end{table}

\subsection{Results}
\subsubsection{Pass Rates}

In \cref{tab:agent-results-pass-rate,tab:agent-results-files,tab:agent-results-tasks}, we examine pass rates across programming languages, task complexity, task categories as well as token efficiency.
Performance stratified by LLM-generated labels from \cref{sec:task classification} are reported in \cref{sec:appendix task clf}.
In addition, \cref{tab:agent-results-retrieval-full} summarizes both file and CST node retrieval accuracy to get a fine-grained view of the agents' capability to navigate the code repository.
If not indicated otherwise, we report the mean pass rate with associated standard error~($\%_{\pm \text{SE}}$), estimated via bootstrap resampling over $n = 2000$ iterations.

\cref{fig:radar plots full,tab:agent-results-pass-rate,tab:agent-results-pass-rate-by-complexity} reveal significant performance variations across programming languages and change types for the three Sonnet 3.5-based agents.
All agents demonstrate their strongest performance in Python~(\qtyrange{20}{24}{\percent}), however these rates remain relatively modest compared to pass rates in SWE-bench~\citep{jimenez2024swebench}.
Performance in Java~(\qtyrange{11}{16}{\percent}) and particularly TypeScript~(\qtyrange{5}{13}{\percent}) shows significant room for improvement.
This performance distribution connects to our complexity analysis but cannot be explained by it alone, as Java exhibits the highest structural complexity~(\SI{66.06}{\percent} mixed changes, 9.81 average node modifications) yet outperforms TypeScript with its lower complexity metrics (\SI{11.39}{\percent} mixed changes, 2.06 modifications).
Stratifying problems by their complexity~(\cref{tab:agent-results-pass-rate-by-complexity}), models perform best on Class Only and Single Class modifications (\qtyrange{25}{40}{\percent}), while struggling significantly with Mixed changes (\qtyrange{8}{15}{\percent}).
Surprisingly, Function Only and Single Function changes also yield relatively low success rates (around \SI{15}{\percent}), despite their typically more contained scope.
This performance distribution cannot be explained by complexity metrics alone as Java outperforms TypeScript despite having significantly higher structural complexity.
These findings suggest that pass rates stem from a complex interplay between task complexity, node change types, and language-specific factors that likely reflect the distribution of programming languages and structural patterns in LLMs' pretraining data.
\begin{table}[b]
    \centering
    \caption{Average pass rates with standard error by task category and average token usage per instance.}
    \small
    \begin{tabular}{cr|rrr|cc}
        \toprule
        & & \multicolumn{3}{c}{Category}& \multicolumn{2}{c}{Tokens~(avg.)} \\
        Agent & Base LLM & Bug Fix & Feature Req. & Refac. & Input & Output \\
        \midrule
        \agentlessPB & Sonnet 3.5 & \meanerror{8.8}{0.71} & \meanerror{5.2}{1.03} & \meanerror{3.2}{2.27}
        & \num{315461} & \num{10900} \\
        \midrule 
        \sweagentPB & Sonnet 3.5 & \meanerror{10.2}{0.76} & \meanerror{9.5}{1.34} & $\mathbf{\meanerror{16.1}{4.70}}$
        & \num{338828} & \num{679}
        \\
        \midrule
        \multirow{5}{*}{\aiderPB} & Sonnet 3.5 
        & $\mathbf{\meanerror{13.8}{0.89}}$ & $\mathbf{\meanerror{15.1}{1.66}}$ & \meanerror{12.9}{4.25}
        & \num{64521} & \num{845} \\
        & Deepseek R1
        & \meanerror{11.7}{0.82} & \meanerror{10.4}{1.40} & $\mathbf{\meanerror{16.1}{4.70}}$ & \num{53440} & \num{2367 } \\
        & Haiku 
        & \meanerror{9.9}{0.75} & \meanerror{9.9}{1.37} & \meanerror{9.7}{3.72} 
        & \num{64067} & \num{1094} \\
        & Mistral-Large 
        & \meanerror{5.7}{0.57} & \meanerror{6.5}{1.14} & \meanerror{4.8}{2.77} 
        & \num{63946} & \num{7319} \\
        & Llama 3.3 70B
        & \meanerror{6.6}{0.62} & \meanerror{4.1}{0.93} & \meanerror{4.8}{2.77}
        & \num{84311} & \num{2002} \\
        & \makecell[r]{DeepSeek-R1-\\Distill-Llama-70B}
        & \meanerror{5.9}{0.58} & \meanerror{3.2}{0.81} & \meanerror{6.5}{3.14} 
        & \num{58028} & \num{2241} \\
        \bottomrule
    \end{tabular}
    \label{tab:agent-results-tasks}
\end{table}

Table~\ref{tab:agent-results-tasks} breaks down performance by task category.
\aiderPB\ achieves the highest average pass rates, while being more token efficient than other agents run with the same base LLM.
Specifically, \aiderPB\ with Sonnet 3.5 utilizes only \SI{19}{\percent} to \SI{20}{\percent} of the input tokens per instance compared to the other agents evaluated in this table.
Table~\ref{tab:agent-results-files} reveals how performance degrades with increasing task complexity, where all methods reach their maximal pass rate on single file edits.
For tasks that require more file edits, a notable performance decrease can be noticed with instances requiring edits of three or more files, success rates start falling below ten percent.
However, it's important to note that without human annotation of test quality, we cannot definitively attribute this drop solely to task complexity. The specificity of the tests may also be a contributing factor.
Lastly, we analyze pass rates grouped by the LLM-based classes introduced in \cref{sec:task classification} in \cref{fig:pass rates by clf full,fig:pass rates by clf full sampled} of the appendix.

\begin{table}[t]
    \centering
    \caption{Performance of different open-source agents on \bench with varying task complexity in terms of files edited. Number of instances in the dataset are in paranthesis.}
    \small
    \begin{tabular}{cr|ccccc}
        \toprule
          & & \multicolumn{5}{c}{Files to be modified}\\
          Agent & Base LLM & 1 \scriptsize($1085$) & 2 \scriptsize($417$) & 3 \scriptsize($200$) & 4 \scriptsize($130$) & 5+ \scriptsize($278$) \\
        \midrule
        \agentlessPB & Sonnet 3.5 & \meanerror{10.8}{0.94} & \meanerror{7.0}{1.24} & \meanerror{4.0}{1.38} & \meanerror{2.3}{1.33} & \meanerror{2.9}{1.02} \\
        \midrule
        \sweagentPB & Sonnet 3.5 & 
        \meanerror{12.5}{1.01} & \meanerror{10.6}{1.49} & \meanerror{5.0}{1.56} & \meanerror{3.8}{1.71} & \meanerror{7.2}{1.56} 
        \\
        \midrule
        \multirow{5}{*}{\aiderPB} & Sonnet 3.5 
        & $\mathbf{\meanerror{17.7}{1.18}}$ & $\mathbf{\meanerror{13.9}{1.72}}$ & $\mathbf{\meanerror{8.0}{1.89}}$ & \meanerror{6.9}{2.27} & $\mathbf{\meanerror{7.9}{1.63}}$ \\
        & Deepseek R1
        & \meanerror{14.9}{1.09} & \meanerror{9.8}{1.45} & \meanerror{6.0}{1.68} & \bfseries $\mathbf{\meanerror{7.7}{2.35}}$ & \meanerror{6.5}{1.51} \\
        & Haiku 
        & \meanerror{12.9}{1.02} & \meanerror{8.6}{1.37} & \meanerror{4.0}{1.38} & \meanerror{6.2}{2.17} & \meanerror{5.8}{1.43} \\
        & Mistral-Large 
        & \meanerror{7.5}{0.79} & \meanerror{5.0}{1.08} & \meanerror{4.5}{1.47} & \meanerror{2.3}{1.33} & \meanerror{3.6}{1.13} \\
        & Llama 3.3 70B
        & \meanerror{8.3}{0.84} & \meanerror{6.5}{1.21} & \meanerror{2.5}{1.11} & \meanerror{0.8}{0.78} & \meanerror{1.4}{0.72} \\
        & \makecell[r]{DeepSeek-R1-\\Distill-Llama-70B}
        & \meanerror{7.5}{0.79} & \meanerror{4.6}{1.02} & \meanerror{2.0}{0.99} & \meanerror{1.5}{1.07} & \meanerror{2.2}{0.88} \\
        \bottomrule
    \end{tabular}
    \label{tab:agent-results-files}
\end{table}

\subsubsection{Retrieval Metrics}
\cref{tab:agent-results-retrieval-full} presents an evaluation of file and node retrieval metrics, demonstrating varying efficacy of different agent-model combinations.
In file retrieval, performance varies significantly across languages, with \sweagentPB\ achieving the highest recall in Java (\SI{51.6}{\percent}), while \aiderPB\ with Sonnet 3.5 excels in precision (\SI{65.1}{\percent}).
For both JavaScript and TypeScript, \aiderPB\ performs significantly better than other agents in both precision and recall.
Notably, while \agentlessPB\ outperforms all other configurations in Python file and node retrieval, its strength is limited to Python instances alone.
Node retrieval results follow similar patterns, with \aiderPB\ leading in Java, JavaScript, and TypeScript, and \agentlessPB\ maintaining superiority in Python tasks.
In general, we observe a significant gap between Python and the other languages.
The highest file retrieval metrics in Python are ahead of the highest metric for any other language by \SI{9.3}{p{.}p{.}} and \SI{12.5}{p{.}p{.}} for recall and precision, respectively.
The same holds for node retrieval, where Python metrics are ahead by \SI{7.7}{p{.}p{.}} and \SI{12.5}{p{.}p{.}} for recall and precision, respectively.
Lastly, we'd like to stress that, as evidenced by the pass rate of \agentlessPB, high retrieval metrics are (most of the time) a necessary but not sufficient condition for high pass rates. 
\begin{table}[h]
    \caption{File and node retrieval metrics for different open-source agents on \bench.}
    \hspace{-0.45cm}
    \begin{minipage}{\textwidth}
    \centering
    \small
    \begin{tabular}{cr|cc|cc|cc|cc}
        \toprule
           & & \multicolumn{8}{c}{ File Retrieval~(\%)} \\
           & & \multicolumn{2}{c}{ Java} & \multicolumn{2}{c}{  JavaScript} & \multicolumn{2}{c}{ TypeScript} & \multicolumn{2}{c}{ Python} \\
        Agent &  Base LLM & \scriptsize Recall & \scriptsize Precision &
           \scriptsize Recall & \scriptsize Precision & 
           \scriptsize Recall & \scriptsize Precision &
           \scriptsize Recall & \scriptsize Precision \\
        \midrule
         \agentlessPB &  Sonnet 3.5 & 
        \num{29.5} & \num{49.7} &
        \num{23.4} & \num{35.2} &
        \num{17.5} & \num{27.7} &
        \bfseries \num{60.9} & \bfseries  \num{77.6}
        \\
        \midrule
         \sweagentPB &  Sonnet 3.5 & 
        \bfseries \num{51.6} & \num{58.5} &
        \num{27.5} & \num{28.5} &
        \num{29.8} & \num{36.4} &
        \num{59.7} & \num{44.2}
        \\ 
        \midrule
         \multirow{6}{*}{\aiderPB} &  Sonnet 3.5 & 
        \num{41.7} & \bfseries \num{65.1} &
        \bfseries \num{37.1} & \bfseries \num{53.5} &
        \num{33.3} & \bfseries \num{52.0} &
        \num{58.2} & \num{73.8}
        \\
        & Deepseek R1 & 
        \num{37.6} &  \num{53.8} &
        \num{31.5} & \num{40.8} &
        \bfseries \num{33.8} & \num{46.0} &
        \num{54.7} & \num{63.3}
        \\
        &  Haiku & 
        \num{35.0} &  \num{53.2} &
        \num{28.3} & \num{40.3} &
        \num{30.2} & \num{45.1} &
        \num{56.8} & \num{70.3}
        \\
        &  Mistral Large & 
        \num{30.6} &  \num{46.8} &
        \num{21.6} & \num{30.9} &
        \num{25.2} & \num{38.5} &
        \num{47.6} & \num{55.9}
        \\
        &  Llama 3.3 70B & 
        \num{27.7} &  \num{43.3} &
        \num{20.6} & \num{28.9} &
        \num{24.7} & \num{39.6} &
        \num{42.9} & \num{54.5}
        \\
        &  \makecell[r]{DeepSeek-R1-\\Distill-Llama-70B} & 
        \num{31.9} &  \num{47.0} &
        \num{25.0} & \num{31.7} &
        \num{27.1} & \num{36.2} &
        \num{48.7} & \num{59.9}
        \\
        & & \multicolumn{8}{c}{ Node Retrieval~(\%)} \\
         \agentlessPB & Sonnet 3.5 & 
        \num{20.6} & \num{38.9} &
        \num{18.9} & \num{27.6} & 
        \num{17.2} & \num{22.9} &
         \num{38.2} & \bfseries \num{63.6}
        \\
        \midrule
         \sweagentPB &  Sonnet 3.5 & 
        \bfseries \num{32.5} & \bfseries \num{52.3} &
        \num{28.8} & \num{23.7} &
        \num{21.7} & \num{20.6} &
        \bfseries \num{38.6} & \num{61.1}
        \\ 
        \midrule
         \multirow{6}{*}{\aiderPB} &  Sonnet 3.5 & 
         \num{29.2} & \num{51.1} &
         \bfseries \num{30.5} & \bfseries \num{39.9} &
         \num{20.2} & \num{26.7} &
        \num{36.7} & \num{59.9}
        \\
        & Deepseek R1 & 
        \num{24.6} &  \num{40.2} &
        \num{26.2} & \num{31.2} &
        \bfseries \num{23.6} & \bfseries \num{29.4} &
        \num{33.5} & \num{50.9}
        \\
        &  Haiku & 
         \num{22.8} &  \num{38.4} &
         \num{22.8} &  \num{29.1} &
         \num{18.2} &  \num{23.5} &
        \num{33.0} & \num{53.8}
        \\
        &  Mistral Large & 
         \num{24.3} &  \num{21.3} &
         \num{17.5} &  \num{16.4} &
         \num{15.3} &  \num{15.9} &
        \num{38.1} & \num{16.7}
        \\
        &  Llama 3.3 70B & 
         \num{18.7} &  \num{31.5} &
         \num{15.0} &  \num{19.4} &
         \num{15.0} &  \num{18.9} &
        \num{24.4} & \num{39.5}
        \\
        &  \makecell[r]{DeepSeek-R1-\\Distill-Llama-70B} & 
         \num{20.5} &  \num{32.5} &
         \num{19.4} &  \num{22.0} &
         \num{17.8} &  \num{20.5} &
        \num{28.1} & \num{44.9}
        \\
        \bottomrule
    \end{tabular}
    \label{tab:agent-results-retrieval-full}
    \end{minipage}
\end{table}

%% file: content/limitations.tex
While being diverse in terms of languages and tasks, \bench has several limitations that should be considered when interpreting the results and for future work.
\paragraph{Task Diversity} Although \bench provides tasks belonging to three different classes, representing common coding tasks, there is a ``long tail'' of problems that are part of the day-to-day work of software developers that are not addressed in this benchmark.%
We believe that targeting the ``head of the distribution'' of tasks is a good first step, but future work should consider expanding to cover a broader range of software engineering challenges that go beyond code changes and could involve aspects such as application deployment and integration testing.

\paragraph{Evaluation Metrics} Our evaluation metrics, while comprehensive, may not capture all aspects of code quality and correctness.
In particular, our current metrics do not assess adherence to code best practices or repository style guides, maintainability, or the presence of potential security flaws in the generated code.
These aspects are crucial in real-world software development and should be incorporated into future evaluations to provide a more holistic assessment of coding agent performance.

\paragraph{Limits of execution-based evaluation} Execution-based evaluation is
the de-facto standard for coding benchmarks. In fact, the execution of unit 
tests is quick and cheap, providing an abundance of feedback signal.
At the 
same time, the activity of the agent is often constrained and it is not possible
to perform major structural changes in the code. For example, it is not possible
to create a website from scratch or a completely new feature in an application 
since tests collected from the repository may end up evaluating nonexistent 
functions due to a different code structure, not due to a different semantic.
We believe this remains an open area of research and we encourage the community
to reason about issues with the limitations of the benchmarks when discussing 
about the autonomy of software agents.

\paragraph{Verifiability} Another important limitation is the lack of human verification to ensure that all tasks in \bench can actually be solved based on the provided information.
There may be instances where the issue descriptions are ambiguous, incomplete, or potentially unsolvable without additional context, highlighting the need to incorporate human expert verification.
However, it's important to consider that in real-world scenarios, the quality and completeness of issue descriptions vary greatly, and coding agents should be capable of handling this variability.
Future benchmarks should strive to balance the need for verifiable tasks with the preservation of diverse issue description qualities, mirroring the range of scenarios encountered in practical software development.

\paragraph{LLM-based Classifications} For our analysis of the problem statement type and the classification of the description according to their quality, we made extensive use of LLM-based annotations. Our annotations provide complementary information that can guide the development of specialized approaches and adds another dimension to the evaluation which informs about existing gaps. However, LLM-based annotations are not without risk as pointed out in \cite{ahmed2024can} and should be interpreted accordingly.

\paragraph{Data Leakage} As the publicly available data used to create \bench may have been utilized during training of the underlying LLMs, data leakage limits the credibility of benchmarks derived from publicly available data.
The lack of transparency in model training sets, coupled with the increasing frequency of model releases, creates an ever-shrinking window to develop truly novel evaluation data.
This situation complicates the evaluation process and underscores the need for innovative approaches to mitigate data leakage effects.
Future work could explore methods like testset slot guessing \citep{deng2023benchmark} to address these challenges and ensure more reliable evaluations of coding agents.

%% file: content/conclusions.tex
We introduced \bench, a repository-level, multi-language benchmark for execution-based evaluation of coding agents.
\bench comprises 2110 samples from 21 repositories across Java, JavaScript, TypeScript, and Python, covering bug fixes, feature requests, and code refactoring.
We also provided \benchSmall, a stratified subset for efficient experimentation.
Our evaluation of leading open-source coding agents required significant effort to modify these agents to work with multiple languages, highlighting the current limitations in their adaptability.
The evaluation revealed significant performance variations across languages, with agents showing stronger capabilities in Python tasks while being less effective in the other three languages.
We also observed a consistent decline in performance as task complexity increased, particularly for multi-file edits and issues that require modifications in classes \textit{and} functions at the same time.
We introduced syntax tree-based retrieval metrics, providing increased granularity to better understand the agents' capabilities in navigating the code repository.
Our findings underscore the need for more versatile and robust AI coding assistants capable of handling complex real-world software engineering tasks across multiple programming languages.
\bench aims to drive progress in developing such agents by providing a comprehensive, multi-lingual evaluation framework.

%% file: content/appendix.tex
\newpage
\appendix
\section{Appendix}

\subsection{Prompts}
\subsubsection{Prompt for classification of tasks} \label{sec:classification_task}
\begin{lstlisting}
prompt = """
    You will be provided with a problem description provided by a user to a github repository, which is labeled as an issue in github, along with the patch that solves the problem. Your task is to try to classify the problem in to a category from the list of categories provided below.
    Problem description:
    {{problem_statement}}
    The gold patch is a diff file that addresses the changes made to the files in the repository in order to solve the issue. It contains the list of files modified or added or removed and the code lines that have been added or replaced or removed.
    Following is the gold patch in a diff format that solves the issue:
    {{gold_patch}}
    
    This is the list of classes we want to classify into, each with description of which issue would the class as its label:
    "Bug Fix": "the problem asks for addressing bugs or issues reported",
    "Feature": "the problem is about introducing new features or enhancements",
    "Testing": "the problem is about adding new testing methods for given code or refactoring existing tests. These could be unit or integration (e2e) tests",
    "Refactoring": "the problem suggests to refactor existing code without changing its external behavior. This could include improving code readability, performance optimizations, or restructuring code for better maintainability",
    "Security": "the problem asks to address security vulnerabilities or concerns in the codebase",
    
    You should output the selected class in the XML format mentioned below. You should only classify into exactly one class. An example output will look like:
    ```
    <category>Feature</category>
    ```
    You must not include any additional text other than the XML and no additional XML tags as well. The value within XML tags should be exactly the same as one of the categories: Bug Fix, Feature, Refactoring.
    """
\end{lstlisting}

\subsubsection{Prompts for classification of problem statements} \label{sec:prompt_classification_ps}

\begin{lstlisting}
 prompt = """
   Your job is to do the following three things:
    1. You will classify the issues description according to its level of detail. 
    2. You will classify the issues description according to whether it is solvable given the provided information. 
    3. You will classify the issues description according to whether it mentions precise code locations to be changed. 

    For tasks 1 to 3 I will also provide you with the correct solution of the problem, termed ground truth. 

    Here is a detailed description of the tasks:

    ### TASK 1 ###
    You will assess if a github issue description is sufficiently detailed such that a software engineer can implemented the solution for the issue after inspecting the code base.

    You will have access to the issue description as well as a ground truth code patch, that is the desired solution. You will not have access to the code base itself. 

    You will provide a brief explanation for your decision and then provide a label, `A`, `B`, `C` or `D` in the XML tags.
    Here are the different lables that you will use:
    `A` contains enough information in natural language to solve the issue
    `B` contains a reproducible failure example 
    `C` contains a partially reproducible example
    `D` does not contain enough information to solve the issue

    Here is the format of the output: 

    <explanation_description>YOUR_EXPLANATION</explanation_description>
    <label_description>YOUR_LABEL</label_description>

    ### TASK 2 ###

    You will help me evaluating the quality of a github issues description together with a ground truth patch that solves the described problem. 
    In particular, your tasks is to check whether the solution or steps to solve the problem are already provided in the issue description.

    You will have access to the issue description as well as a ground truth code patch, that is the desired solution. You will not have access to the code base itself. 

    You will provide a brief explanation for your decision and then provide a label, `A`, `B`, `C`, `D` or `E` in the XML tags.
    Here are the different lables that you will use:
    `A` no solution or steps provided
    `B` partial solution provided (e.g., some steps in natural language)
    `C` complete solution provided (e.g., complete steps in natural language)
    `D` exact patch provided
    `E` misleading solution or steps

    Here is the format of the output: 

    <explanation_solution>YOUR_EXPLANATION</explanation_solution>
    <label_solution>YOUR_LABEL</label_solution>

    ### TASK 3 ###

    You will help me evaluating the quality of a github issues description together with a ground truth patch that solves the described problem. 
    In particular, your tasks is to check whether the issue description contains information on the issue location, i.e., which part of the code 
    needs to modified or fixed to address the issue.

    You will have access to the issue description as well as a ground truth code patch, that is the desired solution. 
    You will not have access to the code base itself. 

    You will provide a brief explanation for your decision and then provide a label, `A`, `B`, `C`, `D` 
    in XML tags. You have to assign exactly one label per issue description. 
    Here are the different lables that you will use:
    `A` exact locations in natural language provided
    `B` exact locations provided in failure stack traces
    `C` related keywords in the issue description are provided that can be used to search for the location
    `D` no location provided. 

    Here is the format of the output: 

    <explanation_location>YOUR_EXPLANATION</explanation_location>
    <label_location>YOUR_LABEL</label_location>
    """

\end{lstlisting}

\subsection{Details CST retrieval metrics}\label{sec:cst details}

For node-level retrieval, we identify the deepest node of the concrete syntax tree~(CST) accompanying a change.
Let a CST be defined by the tuple $(\nodes, \edges, \nodeLabels, \rootNode, \nodeLabelingFunc, \rangeAssignFunc)$, where \nodes\ is the set of all vertices, $\edges \subseteq \nodes \times \nodes$ the set of directed edges, \nodeLabels\ a finite set of node labels~(e.g., \texttt{class}, \texttt{function}), $\rootNode \in \nodes$ the root node~(e.g., a python module), and \nodeLabelingFunc\ a map $\nodeLabelingFunc: \nodes \rightarrow \nodeLabels$ assigning a label to a node.
Furthermore, let a line span be defined as an interval $\lineSpan:=[s,e]$ where $s, e \in \mathbb{N}^{+}$ and $s < e$.
Lastly, $\rangeAssignFunc: \nodes \rightarrow \lineSpan$ is a map assigning a line span to a node.
Since the line span of a higher-level node encompasses the spans of their descendants~(e.g., a function's span lies in the interval of its parent's class's span), we want to identify the deepest node that is affected by a change.
Formally, let $\change \in \lineSpan$ be the line span of a change on the code base.
All nodes affected by a given change are defined as the set of nodes overlapping with the change~(they have a non-empty overlap):
\[
\text{affected}(c, \text{CST}) := \{v \in \nodes | c \cap \rangeAssignFunc(v) \neq \emptyset\}.
\]
The deepest node affected by change \change\ can then defined as
\[
\text{deepestNode}(\change, \text{CST}) := v \in \text{affected}(\change, \text{CST}) | \nexists u \in \text{affected}(\change, \text{CST}): \rangeAssignFunc(u)  \subset \rangeAssignFunc(v).
\]
In other words, the deepest node is the one among all affected nodes that does not contain another affected node.

To compute node-level retrieval metrics, we obtain $\GTNodes^{i}$ as the set of deepest nodes modified in the ground truth patch of task $i$, and $\predictedNodes^{i}$ as the set of deepest nodes affected by the predicted code changes for the task.
Note, that for simplicity, we omitted indexing CST and changes with a file name.
Naturally, we assume that the CST was constructed for the file in which changes were made.

\subsection{Task classifications and pass rates}\label{sec:appendix task clf}

\begin{figure}[h!]
    \centering
    \includegraphics[width=0.95\textwidth]{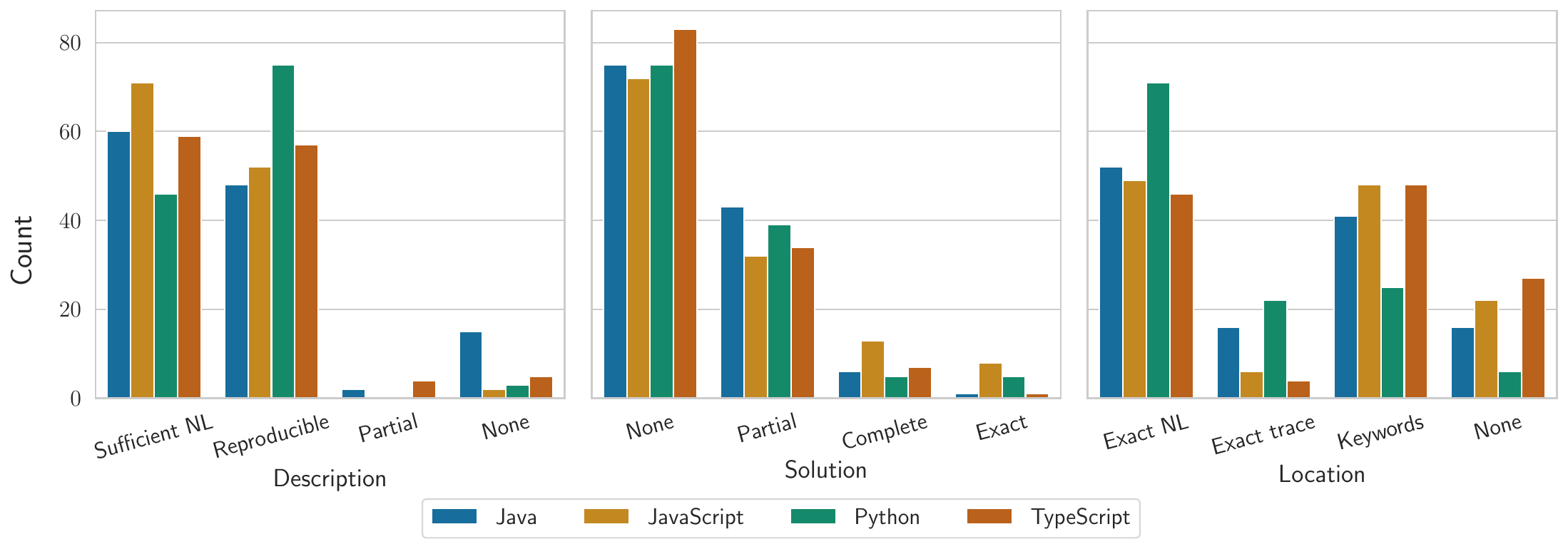}
    \caption{Classification of \benchSmall issue descriptions with respect to their descriptiveness (left plot), hints a the solution (middle plot) and information on the localization of the issue (right plot} \label{fig:categories_ps_sampled}
\end{figure}

Following the categories in \cite{xia2024agentlessdemystifyingllmbasedsoftware} we use the categories in \cref{tab:informativeness_problem_statements} for our classification, roughly ordered with respect to their level of information content. 
\cref{fig:categories_ps_sampled} shows the distribution of task categories on the sub-sampled \benchSmall dataset.

\begin{table}[htb]
\centering
\caption{Informativeness of the problem statements}
\begin{tabular}{ll}
\hline
\textbf{Category} & \textbf{Description of the categories} \\
\hline
\multicolumn{2}{l}{\textit{Descriptiveness of the problem statement}} \\
A & Contains enough information in natural language to solve the issue \\
B & Contains a reproducible failure example \\
C & Contains a partially reproducible example \\
D & Does not contain enough information to solve the issue \\
\hline
\multicolumn{2}{l}{\textit{Solution content already present in the problem statement}} \\
A & No solution or steps provided \\
B & Partial solution provided (e.g., some steps in natural language) \\
C & Complete solution provided (e.g., complete steps in natural language) \\
D & Exact patch provided \\
E & Misleading solution or steps provided \\
\hline
\multicolumn{2}{l}{\textit{Location information on the required changes}} \\
A & Exact locations in natural language provided \\
B & Exact locations provided in failure stack traces \\
C & Related keywords provided that can be used to search for the location \\
D & No location provided \\
\hline
\end{tabular}
\label{tab:informativeness_problem_statements}
\end{table}

\begin{figure}
    \centering
    \begin{subfigure}[b]{\textwidth}
        \centering
        \includegraphics[width=\textwidth]{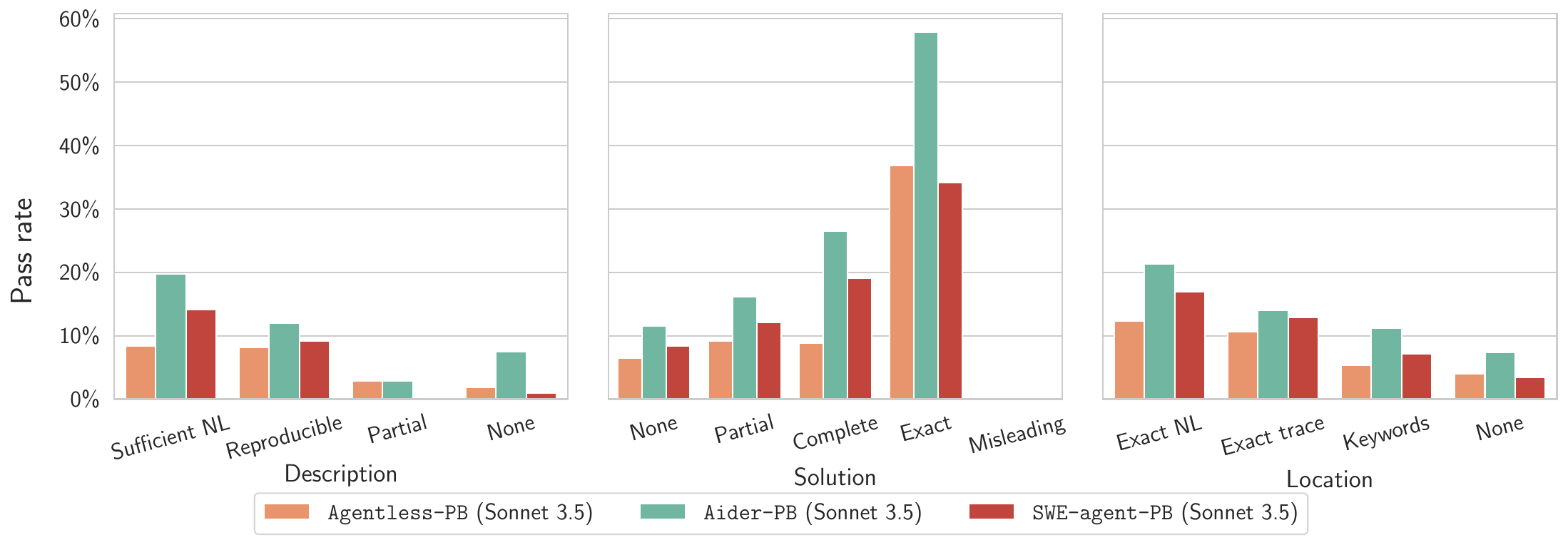}
        \caption{Pass rates by task classification on \bench.}
        \label{fig:pass rates by clf full}
    \end{subfigure}
    \vfill
    \begin{subfigure}[b]{\textwidth}
        \centering
        \includegraphics[width=\textwidth]{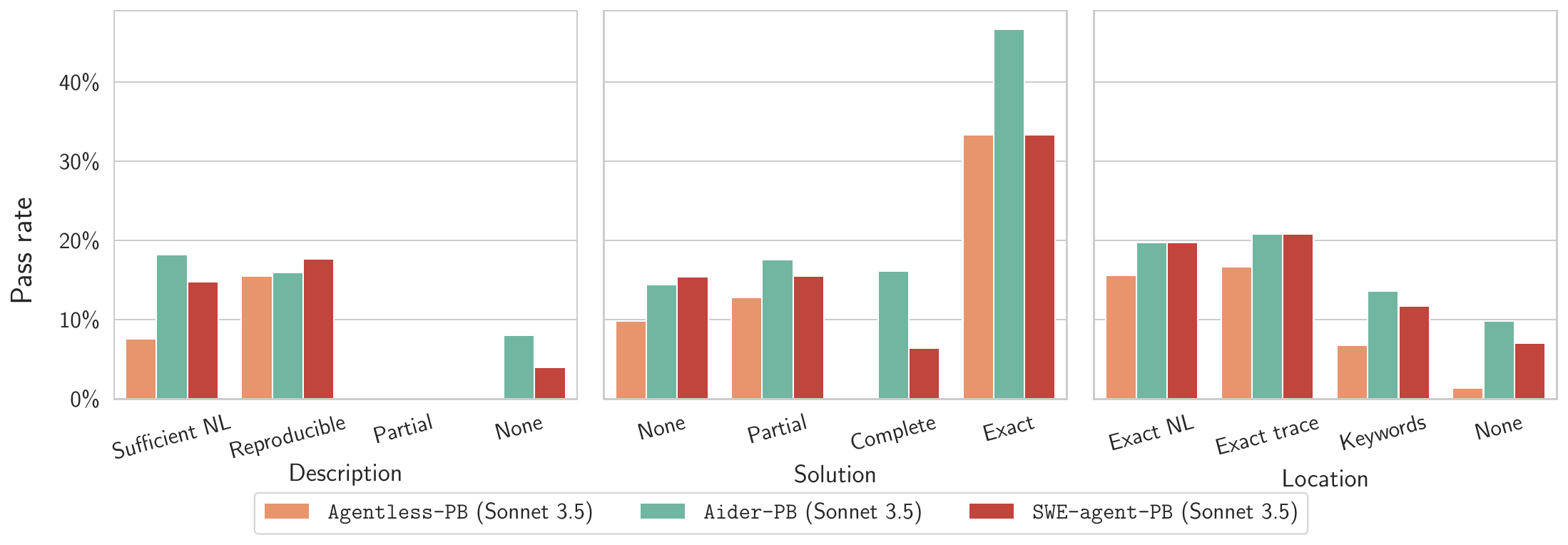}
        \caption{Pass rates by task classification on \benchSmall.}
        \label{fig:pass rates by clf full sampled}
    \end{subfigure}
    \caption{Pass rates of instances with respect to their descriptiveness (left plot), hints a the solution (middle plot) and information on the localization of the issue (right plot).}
    \label{fig:pass rates by clf}
\end{figure}

In \cref{fig:pass rates by clf} we highlight how the level of informativeness of the problem statements impact the pass rates across agents. Overall, more informative problem statements, be it with respect to location, hints at the solution or level of descriptiveness, result in higher pass rates. This confirms the intuition that less details in the problem statement make it more difficult for a task being solved. %

\subsection{Results on \benchSmall}

\begin{figure}[h!]
    \centering
    \includegraphics[width=0.95\linewidth]{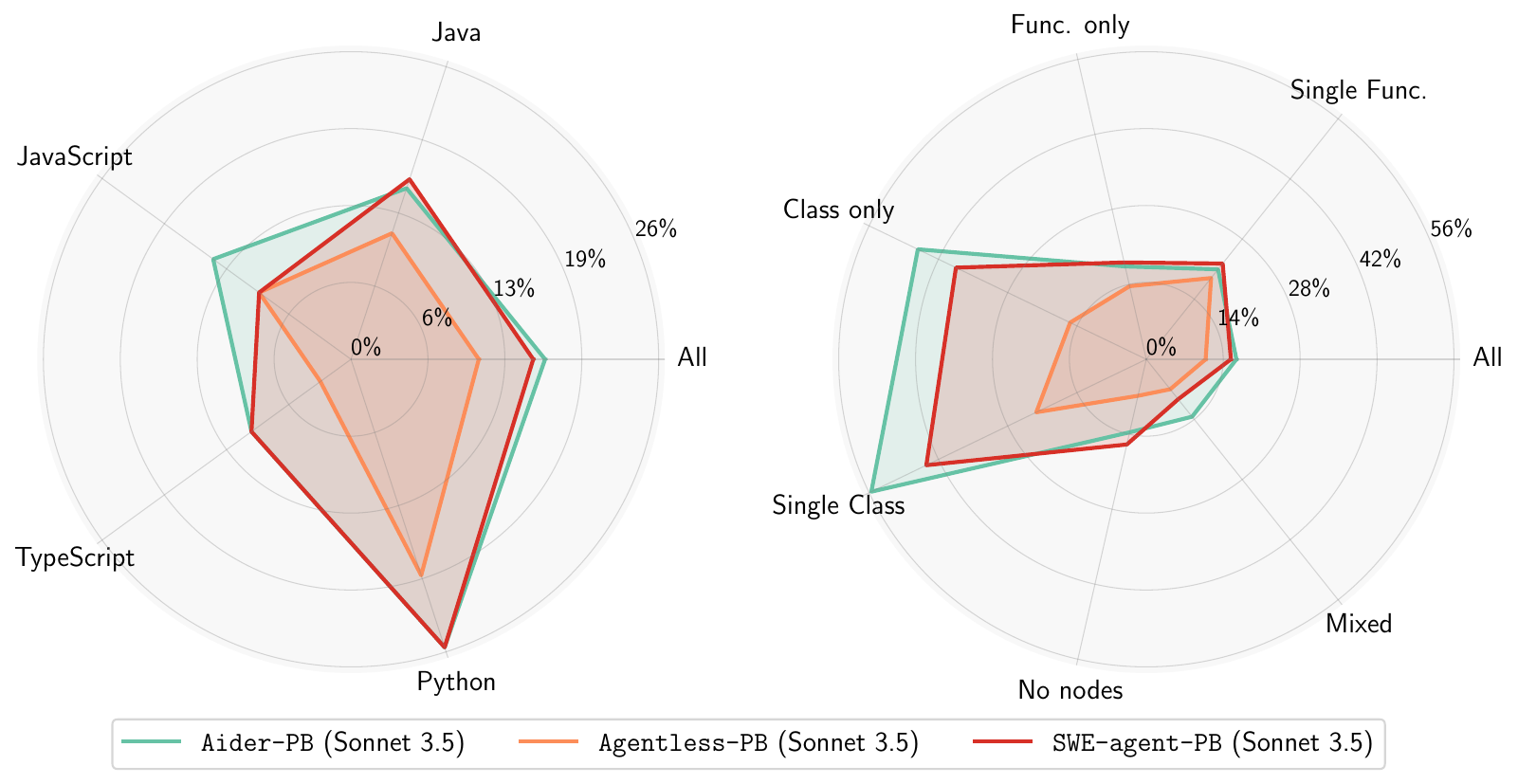}
    \caption{\benchSmall pass rates of coding agents across programming languages~(left) and across subsets of different complexities based on syntax tree nodes. The right plot categorizes changes by type (class or function) and scope (single or multiple), with "No nodes" indicating no class or function changes and "Mixed" requiring both.}
    \label{fig:pass rates sampled}
\end{figure}

\cref{fig:pass rates sampled} presents the coding agents' performance  across different programming languages and code change complexities.
The left radar chart shows the pass rates for Java, JavaScript, TypeScript, Python, and overall performance.
The right chart illustrates the agents' effectiveness in handling various types of code modifications, ranging from changes confined to a single class or function to more complex scenarios involving multiple structural elements.

\begin{table}[t!]
    \centering
    \caption{Pass rates of open source agents on \bench by complexity in terms of CST node changes.}
    \scriptsize
    \begin{tabular}{cr|cccccc}
        \toprule
           & & \multicolumn{6}{c}{Node Change Category}
           \\
           Agent & Base LLM & \makecell{None\\ \tiny($n=263$)} & \makecell{Single Func.\\ \tiny($n=848$)} &  \makecell{Func. Only\\ \tiny($n=1431$)} & \makecell{Single Class\\ \tiny(25)} & \makecell{Class Only\\ \tiny($n=30$)} & \makecell{Mixed\\ \tiny($n=386$)}
           \\
        \midrule
        \agentlessPB
        & Sonnet 3.5 & \meanerror{3.8}{1.20} & \meanerror{11.2}{1.08} & \meanerror{8.7}{0.74} & \meanerror{32.0}{9.42} & \meanerror{26.7}{8.10} & \meanerror{5.7}{1.18} \\
        \midrule
        \sweagentPB
        & Sonnet 3.5 & \meanerror{14.4}{2.23} & \meanerror{11.3}{1.09} & \meanerror{9.6}{0.79} & $\mathbf{\meanerror{52.0}{9.96}}$ & $\mathbf{\meanerror{46.7}{9.12}}$ & \meanerror{6.5}{1.26}
        \\
        \midrule
        \multirow{5}{*}{\aiderPB}
        & Sonnet 3.5 
        & $\mathbf{\meanerror{19.8}{2.48}}$ & $\mathbf{\meanerror{17.0}{1.29}}$ & $\mathbf{\meanerror{13.8}{0.94}}$ & \meanerror{40.0}{9.78} & \meanerror{36.7}{8.93} & $\mathbf{\meanerror{9.3}{1.48}}$
        \\
        & Deepseek R1
        & \meanerror{17.5}{2.36} & \meanerror{13.7}{1.18} & \meanerror{10.8}{0.83} & \meanerror{40.0}{9.78} & \meanerror{36.7}{8.93} & \meanerror{8.3}{1.41} \\
        & Haiku 
        & \meanerror{16.3}{2.30} & \meanerror{12.3}{1.12} & \meanerror{9.3}{0.77} & \meanerror{24.0}{8.58} & \meanerror{23.3}{7.80} & \meanerror{6.5}{1.26}
        \\
        & Mistral-Large 
        & \meanerror{11.8}{2.02} & \meanerror{7.1}{0.88} & \meanerror{5.4}{0.58} & \meanerror{20.0}{8.03} & \meanerror{20.0}{7.41} & \meanerror{2.6}{0.82}
        \\
        & Llama 3.3 70B 
        & \meanerror{11.4}{1.98} & \meanerror{7.4}{0.90} & \meanerror{5.2}{0.58} & \meanerror{32.0}{9.42} & \meanerror{26.7}{8.10} & \meanerror{3.6}{0.96}
        \\
        & \makecell[r]{DeepSeek-R1-\\Distill-Llama-70B} & \meanerror{7.6}{1.63} & \meanerror{7.1}{0.88} & \meanerror{5.0}{0.57} & \meanerror{28.0}{9.00} & \meanerror{26.7}{8.10} & \meanerror{3.1}{0.89}
        \\
        \bottomrule
    \end{tabular}
    \label{tab:agent-results-pass-rate-by-complexity}
\end{table}

\subsection{Collected Repositories} \label{sec:repo_list}

\begin{table}[h!]
    \small
    \centering
    \caption{Contrasting numbers of processed pull requests at the beginning of data collection and at the end as well as average repository size measured in number of files.}

    \begin{tabular}{ccccccc}
        \toprule
        Language   & \makecell{Total \#repos                                                                   \\collected} & \makecell{Total \#PRs\\collected} & \makecell{Total \#PRs\\w/ tests} & \makecell{Final \#\\samples} & \makecell{Final \#\\repos} & \makecell{Avg. repository\\size (files)} \\
        \midrule
        Java       & \num{17}                & \num{69374}  & \num{1433} & \num{165}  & \num{6} & \num{2420.6} \\
        JavaScript & \num{12}                & \num{138713} & \num{3136} & \num{1078} & \num{4} & \num{3706.5} \\
        Python     & \num{10}                & \num{57327}  & \num{1012} & \num{199}  & \num{6} & \num{1928.1} \\
        TypeScript & \num{10}                & \num{151836} & \num{3042} & \num{729}  & \num{5} & \num{8946.0} \\
        \bottomrule
    \end{tabular}
    \label{tab:repo_stats}
\end{table}

\begin{table}[h!]
    \scriptsize
    \centering
    \begin{tabular}{llr}
    \toprule
         Language&  Repository& \#PRs collected\\
         \midrule
         \multirow{17}{*}{Java}&  spring-projects/spring-boot& 6286\\
         &  PhilJay/MPAndroidChart& 378\\
         &  spring-projects/spring-framework& 4792\\
         &  google/guava& 2268\\
         &  NationalSecurityAgency/ghidra& 1044\\
         &  ReactiveX/RxJava& 3906\\
         &  apache/dubbo& 7165\\
         &  skylot/jadx& 536\\
         &  apolloconfig/apollo& 1676\\
 & netty/netty&7524\\
 & Netflix/Hystrix&760\\
 & google/gson&933\\
 & libgdx/libgdx&3576\\
 & apache/rocketmq&3794\\
 & thingsboard/thingsboard&5378\\
 & JetBrains/intellij-community&2518\\
 & trinodb/trino&16840\\
 \midrule
 \multirow{12}{*}{JavaScript}& vercel/next.js&22087\\
 & nodejs/node&33429\\
 & axios/axios&1482\\
 & mrdoob/three.js&16172\\
 & facebook/react&15393\\
 & twbs/bootstrap&15096\\
 & sveltejs/svelte&5324\\
 & atom/atom&5249\\
 & angular/angular.js&7928\\
 & lodash/lodash&1383\\
 & prettier/prettier&9613\\
 & serverless/serverless&5557\\
 \midrule
 \multirow{10}{*}{TypeScript}& freeCodeCamp/freeCodeCamp&36730\\
 & microsoft/vscode&30660\\
 & angular/angular&27565\\
 & mui/material-ui&22533\\
 & puppeteer/puppeteer&5831\\
 & storybookjs/storybook&12461\\
 & tailwindlabs/tailwindcss&2655\\
 & gothinkster/realworld&795\\
 & supabase/supabase&10743\\
 & coder/code-server&1863\\
 \midrule
 \multirow{10}{*}{Python}& Significant-Gravitas/AutoGPT&3939\\
 & huggingface/transformers&16135\\
 & langchain-ai/langchain&13358\\
 & yt-dlp/yt-dlp&2701\\
 & tensorflow/models&3632\\
 & tiangolo/fastapi&3056\\
 & keras-team/keras&7310\\
 & localstack/localstack&5641\\
 & geekan/MetaGPT&773\\
 & 3b1b/manim&782\\
 \bottomrule
    \end{tabular}
    \caption{List of repositories and total number of PRs collected for four languages.}
    \label{tab:repo_list}
\end{table}